\documentclass[english,aps,prb,showpacs,superscriptaddress,floats,amsmath,amssymb,floatfix,nobalancelastpage,twocolumn]{revtex4-2}
\usepackage[T1]{fontenc}
\usepackage[latin9]{inputenc}
\usepackage{babel}
\usepackage{units}
\usepackage{amsmath}
\usepackage{amssymb}
\usepackage{graphicx}
\usepackage[unicode=true]
 {hyperref}

\makeatletter

\newcommand*\LyXThinSpace{\,\hspace{0pt}}
\newcommand{\lyxdot}{.}

\usepackage{accents}

\makeatother

\begin{document}
\title{Finite temperature electric field induced two-dimensional coherent
nonlinear spectroscopy in a Kitaev magnet}
\author{Wolfram Brenig}
\email{w.brenig@tu-bs.de}

\affiliation{Institute for Theoretical Physics, Technical University Braunschweig,
D-38106 Braunschweig, Germany}
\author{Olesia Krupnitska}
\email{olesia.krupnitska@tu-braunschweig.de}

\affiliation{Institute for Theoretical Physics, Technical University Braunschweig,
D-38106 Braunschweig, Germany}
\affiliation{Institute for Condensed Matter Physics, National Academy of Sciences
of Ukraine, Svientsitskii Street 1, 790\LyXThinSpace 11, L'viv, Ukraine}
\begin{abstract}
We study electric field induced two-dimensional coherent nonlinear
optical spectroscopy (2DCS) in a Kitaev magnet at finite temperature.
We show that 2DCS is susceptible to both types of fractional quasiparticles
of this quantum spin-liquid, i.e., fermions and flux visons. Focusing
on the second order response, we find a strong antidiagonal feature
in the two-dimensional frequency plane, related to the galvanoelectric
effect of the fractional fermions. Perpendicular to the antidiagonal,
the width of this feature is set by quasiparticle relaxation rates
beyond the bare Kitaev magnet, thereby providing access to single-particle
characteristics within the multi-particle 2DCS continuum. While the
structure of the 2DCS susceptibility stems from the fermionic quasiparticles
and displays Fermi blocking versus temperature, the emergent bond
randomness which arises due to thermally populated visons strongly
modifies the fermionic spectrum. Therefore also the presence of gauge
excitations is manifest in the 2DCS susceptibility as the temperature
is increased beyond the flux proliferation crossover. Our results
are consistent with and extend previous findings on second harmonic
generation in Kitaev magnets.
\end{abstract}
\maketitle

\section{Introduction}

Two-dimensional (2D) coherent nonlinear optical spectroscopy (2DCS)
\citep{Mukamel2000,Brixner2004,Cho2008} can be viewed as an analog
of 2D nuclear magnetic resonance \citep{Bax1986,Ernst1987,Jeener1971}.
Similar to the latter, and by Fourier transforming the nonlinear response
of a system which is driven by particular sequences of incident ultrashort
laser pulses \citep{Mukamel2000,Brixner2004,Cho2008}, spectral information
can be obtained in more than one frequency dimension. In molecules,
low-dimensional semiconductors, and nanomaterials, this has lead to
a wealth of information about structure, electronic and vibrational
excitations, dynamics, relaxation, and dephasing \citep{Cho2008,Huse2005,Cowan2005,Kuehn2009,Kuehn2011}.

Recently 2DCS has come into focus also to analyze excitations in many-body
phases of correlated electronic and magnetic solid-state systems,
studying magnons \citep{Lu2017}, solitons \citep{Wan2019}, Majorana
fermions and visons \citep{Choi2020,Negahdari2023,Qiang2023}, spinons
\citep{Li2021,Sim2023L,Gao2023,Sim2023,ZLLi2023,Potts2023,Watanabe2024,Kaib2024},
fractons \citep{Nandkishore2021}, marginal Fermi liquids \citep{Mahmood2021},
and three-particle correlations in Anderson and Hubbard models \citep{Kappl2023}.
Several of these studies \citep{Wan2019,Choi2020,Negahdari2023,Qiang2023,Li2021,Sim2023L,Gao2023,Sim2023,ZLLi2023,Potts2023,Watanabe2024,Nandkishore2021}
focus on magnets from the topical research field of quantum spin liquids
(QSL) \citep{Savary2016}. In these magnets, fractionalization induced
nonlocal elementary excitations lead to ubiquitous continua which
cannot be disentangled into their constituents in the spectra of conventional
linear response probes. Here the ability of 2DCS to separate homogeneous
and inhomogeneous broadening \citep{Cho2008,Cundiff2013} has been
highlighted as a promising tool to deconvolute fractionalized continua
\citep{Wan2019}.

The planar Kitaev magnet \citep{Kitaev2006} is among the QSLs of
great current interest \citep{Takagi2019,Motome2019}. It is an Ising
model on the honeycomb lattice with compass anisotropy of its exchange
and allows for fractionalization of spins into mobile Majorana matter
and $\mathbb{Z}_{2}$ gauge flux excitations (visons). The latter
are localized in the absence of external magnetic fields \citep{Kitaev2006}.
The flux-free state can be treated analytically \citep{Kitaev2006},
displaying gapless fermionic quasiparticles. All spin correlations
are short ranged \citep{Baskaran2007}. At nonzero magnetic field,
the spectrum acquires a gap and a chiral edge mode \citep{Kitaev2006}.
Mott-Hubbard insulators with strong spin-orbit coupling \citep{Jackeli2009},
like $\alpha$-RuCl$_{3}$ \citep{Plumb2014}, may display low-energy
magnetic properties close to that of the Kitaev model. However, non-Kitaev
exchange interactions in $\alpha$-RuCl$_{3}$ lead to zigzag antiferromagnetic
order below $7.1$K \citep{Cao2016}. Suppressing this order by in-plane
magnetic fields \citep{Sears2017,Anja2017} opens up a range of $H{\parallel}a\sim7{\dotsc}9$T
which may host a low-temperature QSL \citep{Baek2017,Hentrich2018,Balz2019,Schonemann2020,Balz2021}.
Excitation continua observed in several linear probes, both, direct,
i.e., Raman \citep{Sandilands2015,Nasu2016,Wulferding2020}, inelastic
neutron \citep{Knolle2014,Banerjee2016,Banerjee2017,Do2017}, and
resonant X-ray scattering \citep{Halasz2016}, as well as indirect,
i.e., phonon spectra \citep{Metavitsiadis2020,Metavitsiadis2022,Ye2020,Feng2021,Feng2022,Li_diff_2021},
and ultrasound propagation \citep{Hauspurg2023}, have been attributed
to fractionalization.

Theoretical analysis of 2DCS spectra of Kitaev magnets \citep{Choi2020,Negahdari2023,Qiang2023}
has focused solely on the direct coupling of the driving external
\emph{magnetic} field component of the laser light to the spin and
on zero temperature. While several options for coupling to the \emph{electric}
field component exist in quantum magnets, e.g. \citep{Lorenzana1995,Katsura2005,Jia2007,Katsura2009,Tokura2014,Chari2021},
they remain unexplored for 2DCS in QSLs. Recently, two investigations
\citep{Kanega2021,Krupnitska2023} have made a step into that direction,
considering higher-harmonic generation (HHG) in Kitaev magnets based
on the electric-field induced exchange-striction mechanism \citep{Katsura2009}.
While HHG is also a nonlinear spectroscopy, it lacks the 2D frequency
information of 2DCS. Since both studies \citep{Kanega2021,Krupnitska2023}
find that fingerprints of fractional quasiparticles can be read off
from HHG spectra and, moreover, finite temperature was shown to have
a strong impact \citep{Krupnitska2023}, it seems highly desirable
to extend such analysis into the 2D frequency plane.

In turn, in this paper, we study electric field driven 2DCS in a Kitaev
QSL including the effects of finite temperature and focusing on the
leading order nonlinear contribution. Key findings include an anomalously
singular antidiagonal response in the 2D frequency plane. Perpendicular
to the antidiagonal line, one-particle life-times of the fractional
fermions can be read off within a spectrum that otherwise resembles
a continuum due to fractionalization. Moreover, we observe that visons
also have a strong impact on the global structure of 2DCS, however,
leaving the antidiagonal line-width untouched. The paper is organized
as follows: In Sec. \ref{sec:model} we summarize the model. Sec.
\ref{sec:2DCS-Response-Fun} details our evaluation of the 2DCS susceptibilities
for homogeneous and random gauge states, in Sec. \ref{subsec:LTsusc}
and \ref{subsec:HTsusc}, respectively. Results and discussions are
presented in Secs. \ref{subsec:Preliminaries} - \ref{subsec:Random-flux}.
A summary is given in Sec. \ref{sec:Summary}. To avoid unnecessary
replication, App. \ref{sec:UTrafo} lists some known technicalities
for completeness.

\section{The Model \label{sec:model}}

\begin{figure}[tb]
\centering{}\includegraphics[width=0.65\columnwidth]{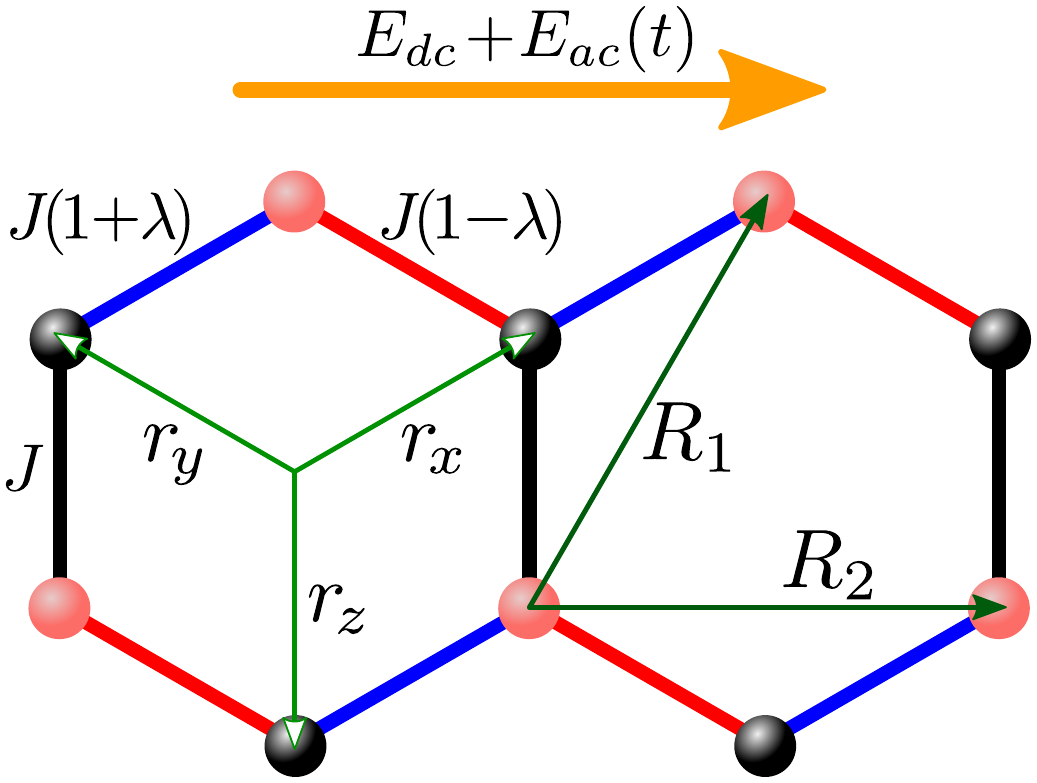}\caption{\label{fig:model} Kitaev model with (blue, red, black) $x$, $y$,
$z$-bonds, hosting $S_{{\bf l}}^{\alpha}S_{{\bf l}+{\bf r}_{\alpha}}^{\alpha}$exchange
with $\alpha{=}x$, $y$, $z$, respectively. ${\bf l}=n_{1}{\bf R}_{1}+n_{2}{\bf R}_{2}$
with basis vectors ${\bf R}_{1[2]}=(1,0),\,[(\frac{1}{2},\frac{\sqrt{3}}{2})]$
and ${\bf r}_{\alpha=x,y,z}=(\frac{1}{2},\frac{1}{2\sqrt{3}}),$ $(-\frac{1}{2},\frac{1}{2\sqrt{3}})$,
$(0,-\frac{1}{\sqrt{3}})$. In electric field $E_{dc}+E_{ac}(t)$
$\perp$ to $z$-bonds, $J\,(1+\lambda,1-\lambda,1$) refers to exchange
interactions on $x$, $y$, $z$-bonds including dimerization $\lambda=-gE_{dc}$
by static field $E_{dc}$.}
\end{figure}

We consider the Kitaev spin-model on the two dimensional honeycomb
lattice \citep{Kitaev2006}
\begin{equation}
H_{0}=\sum_{{\bf l},\alpha}J_{\alpha}S_{{\bf l}}^{\alpha}S_{{\bf l}+{\bf r}_{\alpha}}^{\alpha}\,,\label{eq:H0}
\end{equation}
with Ising exchange $J_{\alpha=x,y,z}$, which we set isotropic, i.e.
$J_{\alpha}=J$ as in Fig. \ref{fig:model}. In the absence of additional
exchange interactions or external magnetic fields, the sign of $J$
is irrelevant. If not denoted explicitly, we use $J$ as the unit
of energy.

The light-matter interaction between the electric field $E$ and the
spin system can be of diverse nature \citep{Tokura2014}. The details
of the 2DCS spectra will depend on that. To make progress, we follow
\citep{Kanega2021} and use a dipole-coupling $-P\cdot E$, based
on the exchange-striction mechanism induced by orbital polarization
\citep{Katsura2009,Tokura2014}
\begin{equation}
P=\frac{\partial H_{0}}{\partial E}=g\sum_{{\bf l}}(S_{{\bf l}}^{x}S_{{\bf l}+{\bf r}_{x}}^{x}-S_{{\bf l}}^{y}S_{{\bf l}+{\bf r}_{y}}^{y})\,.\label{eq:P}
\end{equation}
We use an electric field $\mathbf{E}=E\mathbf{e}_{\perp,z}$ perpendicular
to the $z$-bonds. $P$ is the effective polarization operator and
$g$ is the magnetoelectric coupling constant. Before proceeding,
we explicitly caution that strictly speaking finite $g$ requires
broken inversion symmetry \citep{Katsura2009}. I.e., currently existing
approximate Kitaev systems may realize neither Eq. (\ref{eq:H0})
nor (\ref{eq:P}), even though indications of ferroelectricity have
been reported for $\alpha$-RuCl$_{3}$ \citep{Mi2022} and Na$_{2}$Co$_{2}$TeO$_{6}$
\citep{Mukherjee2022}. Despite these remarks, combining Eq. (\ref{eq:H0})
and (\ref{eq:P}) provides for a legitimate case study of electric
field induced 2DCS on a QSL, which is what we focus on. For Mott-insulators
it has been suggested that $|g/J|\sim O(0.1)$cm/MV can be reached
\citep{Kanega2021}.

Nonlinear susceptibilities of the polarization $P$ at order $N$
of $E$ involve thermal expectation values which are rank-$(N{+}1)$
tensors of $P$ \citep{Butcher1990}. Since the equilibrium density
matrix obeys the symmetries of the Kitaev model, it is invariant under
the operation $U$ of simultaneous reflection on the $z$-bond $(x{,}y){\rightarrow}({-}x{,}y)$,
including an exchange of spins $S^{x{,}y{,}z}{\rightarrow}(+{,}-{,}+)S^{y{,}x{,}z}$.
The polarization and electric field both change sign under this operation.
In turn, even-$N$ susceptibilities vanish unless the $U$-symmetry
of $H_{0}$ is broken.

Since $N=2$ marks not only the lowest order nonlinear response, but
also the leading order 2DCS susceptibility, it proves very useful
that one can explicitly break the $U$-symmetry by retaining the \emph{static},
i.e., DC component of the electric field. This approach is well known
from field-induced second harmonic generation in semiconductors \citep{Aktsipetrov1996}
or graphene \citep{Bykov2012} and it has been applied analogously
for HHG generation in the Kitaev magnet \citep{Kanega2021,Krupnitska2023}.
To see this, we decompose $E=E_{dc}+E_{ac}(t)$ into a static (DC)
and a dynamic (AC) part, the latter of which time-averages to zero.
As shown in Fig. \ref{fig:model}, $E_{dc}$ can be absorbed into
a rescaled exchange $J_{\alpha}=J(1+\lambda,1-\lambda,1)$ with $\lambda=-gE_{dc}$.
This dimerization breaks the $U$-symmetry.

Using the standard mapping to Majorana fermions and a static $\mathbb{Z}_{2}$
gauge field $\eta_{{\bf l}}=\allowbreak\pm1$, residing on, e.g.,
the $\alpha=\allowbreak z$ bonds \citep{Kitaev2006,Motome2019},
the Kitaev magnet with applied electric field reads
\begin{align}
\lefteqn{H_{0}-P(E_{dc}+E_{ac}(t))=H-PE_{ac}(t)=}\label{eq:Ht}\\
 & -\frac{i}{2}\!\sum_{{\bf l},\alpha=x,y,z}\!\!\!\!J_{\alpha}\eta_{{\bf l},\alpha}\,a_{{\bf l}}c_{{\bf l}+{\bf r}_{\alpha}}+\frac{i}{2}\!\sum_{{\bf l},\alpha=x,y}\!\!\!\!\mathrm{sg}_{\alpha}\,a_{{\bf l}}c_{{\bf l}+{\bf r}_{\alpha}}\,gE_{ac}(t)\,,\nonumber 
\end{align}
where $\eta_{{\bf l},x(y)}=1$, $\eta_{{\bf l},z}=\eta_{{\bf l}}$,
and $\mathrm{sg}_{\alpha}=+(-)$ for $\alpha=x(y)$. We choose to
normalize the two Majorana fermions per unit cell according to $\{a_{{\bf l}},a_{{\bf l}'}\}=\delta_{{\bf l},{\bf l}'}$,
$\{c_{{\bf m}},c_{{\bf m}'}\}=\delta_{{\bf m},{\bf m}'}$, and $\{a_{{\bf l}},c_{{\bf m}}\}=0$.

The expression (\ref{eq:Ht}) for the total Hamiltonian explicitly
displays an additional reason for studying the optical exchange-striction
coupling, namely, the Hamiltonian remains diagonal in the gauge flux,
or speaking differently, this type of coupling to light does not excite
visons. In turn, visons will affect the 2DCS spectrum solely through
their thermal occupation.

\begin{figure}[tb]
\centering{}\includegraphics[width=0.7\columnwidth]{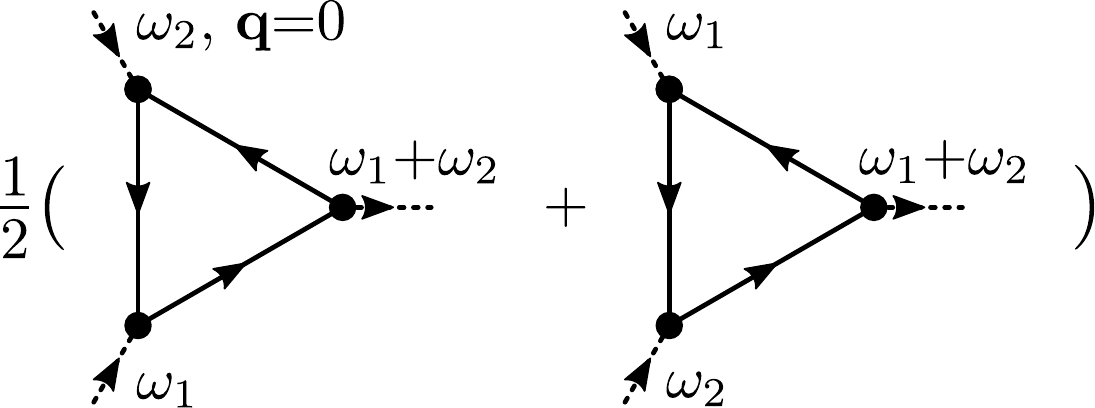}\caption{\label{fig:diags} Diagrams for the 2DCS susceptibility $\chi_{2}(\omega_{1},\omega_{2})$
at order $O(2)$ in $E_{ac}(t)$. The solid lines carry a band index
$\mu=1,2$, the dots refer to the $2\times2$ polarization operator
matrices $p_{\mu\nu}$. I.e., the diagrams correspond to 2$\times$8
expressions.}
\end{figure}

\section{Second order two-dimensional response function\label{sec:2DCS-Response-Fun}}

Technically, we are interested in the Fourier transform into the 2D
frequency plane of the polarization response $\langle\Delta P\rangle(t)$
at $O(N=2)$ in $E_{ac}(t)$, i.e., the retarded susceptibility $\tilde{\chi}_{2}(t,t_{1},t_{2})=i^{2}\Theta(t-t_{1})\Theta(t_{1}-t_{2})\allowbreak\langle[[P(t),\allowbreak P(t_{1})],P(t_{2})]\rangle$
\citep{Krupnitska2023,Butcher1990}. This can be obtained from analytic
continuation to the real axis of the Matsubara frequency transform
of the fully connected contractions of the imaginary time propagator
$\tilde{\chi}_{2}(\tau_{2},\tau_{1})=\langle T_{\tau}(\allowbreak P(\tau_{2})\allowbreak P(\tau_{1})P)\rangle$
\citep{Parker2019,Rostami}, i.e., the standard Feynman diagrams of
Fig. \ref{fig:diags}. It should be noted that the $N$-fold time
integrations in the perturbative expansion of the time-dependent density
matrix at any order $N$ in $P\,E_{ac}(t)$ are totally symmetric
with respect to any permutation of the $N$ time arguments, dubbed
intrinsic permutation symmetry \citep{Butcher1990}. Therefore, any
$N$-th order contributions to $\langle\Delta P\rangle(t)$ can be
accounted for by the fully symmetrized susceptibility $\chi_{N}(t,t_{1},\allowbreak\dots,t_{n})=\allowbreak\sum_{\pi}\tilde{\chi}_{N}(t,t_{\pi(1)},\allowbreak\dots,t_{\pi(N)})/N!$
\citep{Rostami}, where $\pi$ labels all permutations. This corresponds
to the two diagrams of Fig. \ref{fig:diags}.

To evaluate Fig. \ref{fig:diags}, we consider two temperature regimes
in this work, namely, above and below the flux proliferation temperature
$T^{\star}$. It is well established \citep{Motome2019,Nasu2015,Metavitsiadis2017,Pidatella2019}
that in a narrow range of $0.01\lesssim T/J\lesssim0.05$ centered
at $T^{\star}$ the $\mathbb{Z}_{2}$ flux gets thermally populated,
rapidly changing the average gauge link density from $\langle\eta_{\mathbf{l}}\rangle=1$
to $\langle\eta_{\mathbf{l}}\rangle=0$. This allows to treat the
Kitaev model at almost any temperature by considering a \emph{homogeneous}
flux state in Sec. \ref{subsec:LTsusc} for $T\lesssim T^{\star}$
and a \emph{random} flux state \citep{RndSec} for $T\gtrsim T^{\star}$
in Sec. \ref{subsec:HTsusc}. This approach has proven to work well
in studies of thermal conductivity \citep{Metavitsiadis2017,Pidatella2019,Metavitsiadis2017a},
phonon renormalization \citep{Metavitsiadis2020,Metavitsiadis2022},
and second harmonic generation \citep{Krupnitska2023}, including
almost quantitative agreement with exact diagonalization \citep{Metavitsiadis2017}
and quantum Monte-Carlo \citep{Feng2021,Feng2022} where available.
Finally, for the range of $0.01\lesssim T/J\lesssim0.05$, the impact
of flux proliferation can be analyzed phenomenologically by considering
a random flux state, however deliberately varying the average density
of flipped links in the range of $[0,\nicefrac{1}{2}]$ while fixing
the fermion temperature at $T\sim O(T^{\star})$.

\subsection{Homogeneous flux state $T\lesssim T^{\star}$\label{subsec:LTsusc}}

In the homogeneous flux ground-state, i.e., for $\eta_{{\bf l}}=1$,
Hamiltonian (\ref{eq:H0}) can be diagonalized analytically in terms
of complex Dirac fermions. Since this procedure is well documented,
e.g., \citep{Motome2019,Metavitsiadis2020,Metavitsiadis2022,Metavitsiadis2021}
and many refs. therein, we merely quote the notations necessary to
state our results. The Majorana fermions are mapped onto Dirac fermions
on half of the momentum space by Fourier transforming $a(c)_{{\bf k}}^{\phantom{\dagger}}=\allowbreak\sum_{{\bf l}}e^{-i{\bf k}\cdot{\bf l}}\allowbreak a(c)_{{\bf l}}/\allowbreak\sqrt{N}$
with momentum ${\bf k}$. They satisfy $a(c)_{{\bf k}}^{\dagger}=\allowbreak a(c)_{-{\bf k}}^{\phantom{\dagger}}$.
Fermionic anticommutation relations apply, $\{a_{{\bf k}}^{\phantom{\dagger}},\allowbreak a_{{\bf k}'}^{\dagger}\}=\allowbreak\delta_{{\bf k},{\bf k}'}$,
$\{c_{{\bf k}}^{\phantom{\dagger}},\allowbreak c_{{\bf k}'}^{\dagger}\}=\allowbreak\delta_{{\bf k},{\bf k}'}$,
and $\{a_{{\bf k}}^{(\dagger)},\allowbreak c_{{\bf k}'}^{(\dagger)}\}=0$.
The diagonal form of $H$ is
\begin{equation}
H=\tilde{\sum_{{\bf k},\gamma=1,2}}\mathrm{sg}_{\gamma}\,\epsilon_{{\bf k}}\,d_{\gamma{\bf k}}^{\dagger}d_{\gamma{\bf k}}^{\phantom{\dagger}}\,,\label{eq:Hd}
\end{equation}
where $[c_{{\bf k}},a_{{\bf k}}]^{T}=\mathbf{u}(\mathbf{k})\,[d_{1{\bf k}},d_{2{\bf k}}]^{T}$
defines the quasiparticle fermions $d_{\gamma{\bf k}}$ via a unitary
transformation $\mathbf{u}(\mathbf{k})$, listed in App. \ref{sec:UTrafo},
and $\mathrm{sg}_{\gamma}$=1(-1) for $\gamma$=1(2). The quasiparticles
satisfy $\smash{d_{1(2){\bf k}}^{\dagger}=d_{2(1)-{\bf k}}^{\phantom{\dagger}}}$,
and $\smash{\tilde{\sum}}$ sums over half of momentum space. In Cartesian
coordinates the quasiparticle energy $\epsilon_{{\bf k}}$ reads $\epsilon_{{\bf k}}=\allowbreak J[3+\allowbreak2\lambda^{2}+\allowbreak2(1-\allowbreak\lambda^{2})\cos(k_{x})+\allowbreak4\cos(k_{x}/2)\allowbreak\cos(\sqrt{3}k_{y}/2)-\allowbreak4\lambda\sin(k_{x}/2)\allowbreak\sin(\sqrt{3}k_{y}/2)\allowbreak]^{1/2}/2$.
Similar to the Hamiltonian, the polarization can also be transformed
into the diagonal Dirac fermion basis
\begin{equation}
P=g\tilde{\sum_{{\bf k},\mu\nu}}d_{\mu{\bf k}}^{\dagger}p_{\mu\nu}(\mathbf{k})d_{\nu{\bf k}}^{\phantom{\dagger}}\,,\label{eq:Pd}
\end{equation}
where, in Cartesian coordinates, $p_{11}(\mathbf{k})=\allowbreak-p_{22}(\mathbf{k})=\allowbreak\sin(\allowbreak k_{x}/\allowbreak2)\allowbreak(2\lambda\sin(k_{x}/2)-\allowbreak\sin(\sqrt{3}k_{y}/2))/\allowbreak(2\epsilon_{{\bf k}})$
and $p_{12}(\mathbf{k})=\allowbreak p_{21}^{\star}(\mathbf{k})=\allowbreak-i\sin(k_{x}/2)\allowbreak(2\cos(\allowbreak k_{x}/\allowbreak2)+\allowbreak\cos(\allowbreak\sqrt{3}k_{y}/2))/\allowbreak(2\epsilon_{{\bf k}})$.
Evidently, $P$ comprises inter- and intraband transitions.

Using Fig. \ref{fig:diags}, and Eqs. (\ref{eq:Hd}) and (\ref{eq:Pd}),
it is now straightforward to evaluate the 2DCS susceptibility. We
obtain
\begin{align}
\chi_{2}(z_{1},z_{2})=g^{3}\tilde{\sum_{\mathbf{k}}} & \left[\frac{8(1{-}2f_{\mathbf{k}})\epsilon_{\mathbf{k}}^{2}\,p_{11}(\mathbf{k})|p_{12}(\mathbf{k})|^{2}}{(z_{1}^{2}-4\epsilon_{\mathbf{k}}^{2})(z_{2}^{2}-4\epsilon_{\mathbf{k}}^{2})}\right.\nonumber \\
 & \times\left.\frac{(z_{1}^{2}{+}z_{1}z_{2}{+}z_{2}^{2}-12\epsilon_{\mathbf{k}}^{2})}{((z_{1}{+}z_{2})^{2}-4\epsilon_{\mathbf{k}}^{2})}\right]\,,\label{eq:c2s}
\end{align}
where $z_{a}=\omega_{a}{+}i\eta$ with $\eta\rightarrow0^{+}$ and
the Fermi function $f_{\mathbf{k}}=1/(e^{\epsilon_{\mathbf{k}}/T}+1)$.

\subsection{Random flux state $T\gtrsim T^{\star}$\label{subsec:HTsusc}}

For an arbitrary real-space distribution of $\{\eta_{\mathbf{l}}\}$,
the Majorana fermions on the $2N$ sites of the lattice are represented
as a spinor $A_{\text{\ensuremath{\sigma}}}^{\dagger}=(a_{1}\dots\allowbreak a_{{\bf l}}\dots\allowbreak a_{N},\allowbreak c_{1}\dots\allowbreak c_{{\bf l}+{\bf r}_{x}}\dots c_{N}):=\mathbf{A}^{\dagger}$.
Using the unitary Fourier-type transform ${\bf F}$, constructed from
two disjoint $N\times N$ blocks $F_{\sigma\rho}^{i=1,2}=e^{-i{\bf k}_{\sigma}\cdot{\bf R}_{\rho}^{i}}/\sqrt{N}$,
with $\sigma,\rho=1\dotsc N$ and ${\bf R}_{\rho}^{i}={\bf l}$ and
${\bf l}+{\bf r}_{x}$, for $a$- and $c$-Majorana lattice sites,
respectively, we map $\mathbf{A}$ onto Dirac fermions by ${\bf D}={\bf F}{\bf A}$.
The Brillouin zone is divided into pairs of $\pm\mathbf{k}$, with
all ${\bf k}\neq-{\bf k}$ and in practice ${\bf F}$ is rearranged
such as to associate the $d_{1}^{\dagger}\dots d_{N}^{\dagger}$ with
the $2\,(N/2)=N$ 'positive' ${\bf k}$-vectors. We then write
\begin{equation}
H={\bf D}^{\dagger}{\bf h}\,{\bf D}/2\,,\hphantom{aaa}P=g\,{\bf D}^{\dagger}{\bf p}\,{\bf D}/2\,,\hphantom{aaa}\label{eq:HPmat}
\end{equation}
for a given $\{\eta_{\mathbf{l}}\}$, with $\mathbf{h}$ and $\mathbf{p}$
being $2N\times2N$ matrices, in general nondiagonal and particle
number nonconserving. Next, $\mathbf{h}$ is diagonalized by a numerical
Bogoliubov transformation $\mathbf{U}$ onto quasiparticles ${\bf S}=\allowbreak{\bf U}{\bf D}$,
for which $H=\sum_{\rho=1}^{2N}\epsilon_{\rho}S_{\rho}^{\dagger}S_{\rho}^{\phantom{\dagger}}/2$,
with $({\bf U}\mathbf{h}{\bf U}^{\dagger})_{\rho\sigma}=\delta_{\rho\sigma}\epsilon_{\rho}$
and $\epsilon_{\rho}=(\epsilon_{1}\dots\epsilon_{N},-\epsilon_{1}\dots-\epsilon_{N})$.
For the diagram calculations, we stay within the Nambu space of $\rho=1\dotsc2N$,
keeping the particle- and hole-range of $S_{\rho}^{(\dagger)}$.

The quasiparticle Green's function $G_{\alpha\beta}(\tau)=-\langle T_{\tau}(S_{\alpha}^{\phantom{\dagger}}\allowbreak S_{\beta}^{\dagger})\rangle$
in Matsubara frequency space reads $G_{\alpha\beta}(\varepsilon_{n})=\delta_{\alpha\beta}/(i\varepsilon_{n}-\epsilon_{\alpha})$
with $\varepsilon_{n}=(2n+1)\pi T$. It satisfies the following helpful
identities $-\langle T_{\tau}(S_{\alpha}^{\text{\ensuremath{\dagger}}}S_{\beta}^{\dagger})\rangle=-\langle T_{\tau}(S_{\bar{\alpha}}^{\phantom{\dagger}}S_{\beta}^{\dagger})\rangle=G_{\bar{\alpha}\beta}(\tau)$
for the anomalous Green's functions, using the notation $\bar{\rho}=\rho\mp N$
for $\rho\gtrless N$. Equipped with this, and Fig. \ref{fig:diags},
and after some algebra we arrive at the 2DCS susceptibility
\begin{align}
\lefteqn{\text{\ensuremath{\chi_{2}}(\ensuremath{z_{1}},\ensuremath{z_{2}})}=g^{3}\sum_{\alpha\beta\gamma}\frac{t_{\alpha\gamma}t_{\gamma\beta}t_{\beta\alpha}}{2(z_{1}+z_{2}-\epsilon_{\alpha}+\epsilon_{\gamma})}\left[\vphantom{\frac{1}{z_{2}}}\right.}\nonumber \\
 & \hphantom{aaa}(f_{\alpha}-f_{\beta})\left(\frac{1}{z_{2}-\epsilon_{\alpha}+\epsilon_{\beta}}+\frac{1}{z_{1}-\epsilon_{\alpha}+\epsilon_{\beta}}\right)+\label{eq:c2srnd}\\
 & \left.\hphantom{aaa}\left(f_{\gamma}-f_{\beta}\right)\left(\frac{1}{z_{2}-\epsilon_{\beta}+\epsilon_{\gamma}}+\frac{1}{z_{1}-\epsilon_{\beta}+\epsilon_{\gamma}}\right)\right]\,,\nonumber 
\end{align}
for a fixed set of $\{\eta_{\mathbf{l}}\}$, with Fermi function $f_{\alpha}=1/(\exp(\epsilon_{\alpha}/T)+1)$
and $t_{\alpha\beta}=(m_{\alpha\beta}-m_{\bar{\beta}\bar{\alpha}})/2$
from the Bogoliubov transform of the polarization $P=g\,{\bf S}^{\dagger}{\bf m}\,{\bf S}/2$
which is not simultaneously diagonal with $H$, i.e., remains particle-number
nonconserving. Finally and following ref. \citep{Metavitsiadis2017},
$\mathbf{U}$, $\epsilon_{\alpha}$, $m_{\alpha\beta}$, and Eqs.
(\ref{eq:c2srnd}) are calculated numerically for a sufficiently large
number of random distributions $\{\eta_{{\bf l}}\}$ and averaged
over.

Eqs. (\ref{eq:c2s}) and (\ref{eq:c2srnd}) constitute the central
formulas of this paper. Next, we discuss them from various perspectives.

\section{Discussion\label{sec:Discussion}}

\subsection{Preliminaries\label{subsec:Preliminaries}}

We begin the discussion with some consistency checks of Eqs. (\ref{eq:c2s})
and (\ref{eq:c2srnd}). First, in the thermodynamic limit and for
any general distribution $\{\eta_{\mathbf{l}}\}$, there is no obvious
algebraic relation between these two equations. However, one may consider
the exemplary case of a single Bogoliubov-pair of states only in Eq.
(\ref{eq:c2srnd}), with $\alpha=1,2$ and $\epsilon_{1,2}=\epsilon,-\epsilon$.
Carrying out the sum on $\alpha,\beta,\gamma$ for that case, it is
reassuring to realize that Eq. (\ref{eq:c2srnd}) then is identical
to the contribution from a single $\mathbf{k}$-point in Eq. (\ref{eq:c2s}),
identifying $\epsilon_{1,2}=\epsilon_{\mathbf{k}},-\epsilon_{\mathbf{k}}$
and $t_{\alpha\beta}=p_{\alpha\beta}(\mathbf{k})$.

Next, setting $z_{1}=z_{2}=z$ the 2DCS susceptibilities describe
the physics of second harmonic generation. This has been analyzed
in ref. \citep{Krupnitska2023}. Therefore, we mention, that inserting
this case into Eqs. (\ref{eq:c2s}) and (\ref{eq:c2srnd}), one indeed
recovers Eqs. (7) and (12) of the above-mentioned work \citep{Krupnitska2023,GlobSign}.

Now, and for a more direct understanding of Eq. (\ref{eq:c2s}), we
step back and motivate $\chi_{2}(z_{1},z_{2})$ at $T=0$ by simple
consideration of a two-level system. Since the homogeneous state is
translationally invariant, and the wave vector of the laser light
is practically zero, all excitations occur on a disjoint collection
of pairs of states $\{1\mathbf{k},2\mathbf{k}\}$ with energies $\{\epsilon_{\mathbf{k}},-\epsilon_{\mathbf{k}}\}$,
enumerated by $\mathbf{k}$. In principle the levels in each two-level
system refer to a Dirac fermion in either the upper or the lower band,
however at $T=0$ one may focus on a two-dimensional Hilbert space
with a complete orthonormal set of eigenstates of the Hamiltonian
$|1\rangle,\epsilon$ and $|2\rangle,-\epsilon$. I.e., $|1\rangle(|2\rangle)$
refers to the Dirac fermion being in the upper(lower) band. The Hamiltonian
reads $H=\epsilon|1\rangle\langle1|-\allowbreak\epsilon|2\rangle\langle2|$
and in the interaction picture the $2\times2$ polarization operator
is assumed to be $P=n|1\rangle\langle1|-\allowbreak n|2\rangle\langle2|+\allowbreak me^{2i\epsilon t}|1\rangle\langle2|+\allowbreak m^{\star}e^{-2i\epsilon t}\allowbreak|2\rangle\langle1|$.
Note that $P$ is consistent with Eq. (\ref{eq:Pd}) regarding the
sign structure of the diagonal elements. For $\tilde{\chi}_{2}(t,t_{1},t_{2})$
at $T=0$, the matrix element $c(t,t_{1},t_{2})=i^{2}\langle2|[[P(t),P(t_{1})],P(t_{2})]|2\rangle$
needs to be considered \citep{ideqm}
\begin{equation}
c(t{,}t_{1}{,}t_{2})=4n|m|^{2}(\cos(2\epsilon(t_{2}{-}t_{1})){-}\cos(2\epsilon(t_{2}{-}t)))\,.\label{eq:tlc}
\end{equation}
Next, we allow for two driving frequencies of the electric field $E(t)$,
i.e., $e^{-iz_{1}t}$ and $e^{-iz_{2}t}$ with $z_{a}=\omega_{a}+i\eta$,
$\eta\rightarrow0^{+}$, and perform the time ordered, elementary
integrals over $c(t,t_{1},t_{2})$ to obtain the response $\langle\Delta P\rangle(t)$
as, e.g., in \citep{Krupnitska2023,Butcher1990,Rostami}. Finally
we symmetrize over the permutations of $z_{1},z_{2}$ as in Fig. \ref{fig:diags},
and get
\begin{equation}
\chi_{2}(z_{1}{,}z_{2})=\frac{8\epsilon^{2}n|m|^{2}(z_{1}^{2}{+}z_{1}z_{2}{+}z_{2}^{2}-12\epsilon^{2})}{(z_{1}^{2}-4\epsilon^{2})(z_{2}^{2}-4\epsilon^{2})((z_{1}{+}z_{2})^{2}-4\epsilon^{2})}\,.\label{eq:tlchi}
\end{equation}
Satisfyingly, this is identical to Eq. (\ref{eq:c2s}) except for
reintroducing $\mathbf{k}$, the occupational prefactor $(1-2f_{\mathbf{k}})$,
the replacement $n|m|^{2}\rightarrow g^{3}p_{11}(\mathbf{k})|p_{12}(\mathbf{k})|^{2}$,
and finally integrating over $\mathbf{k}$.

\begin{figure}[tb]
\centering{}\includegraphics[width=0.82\columnwidth]{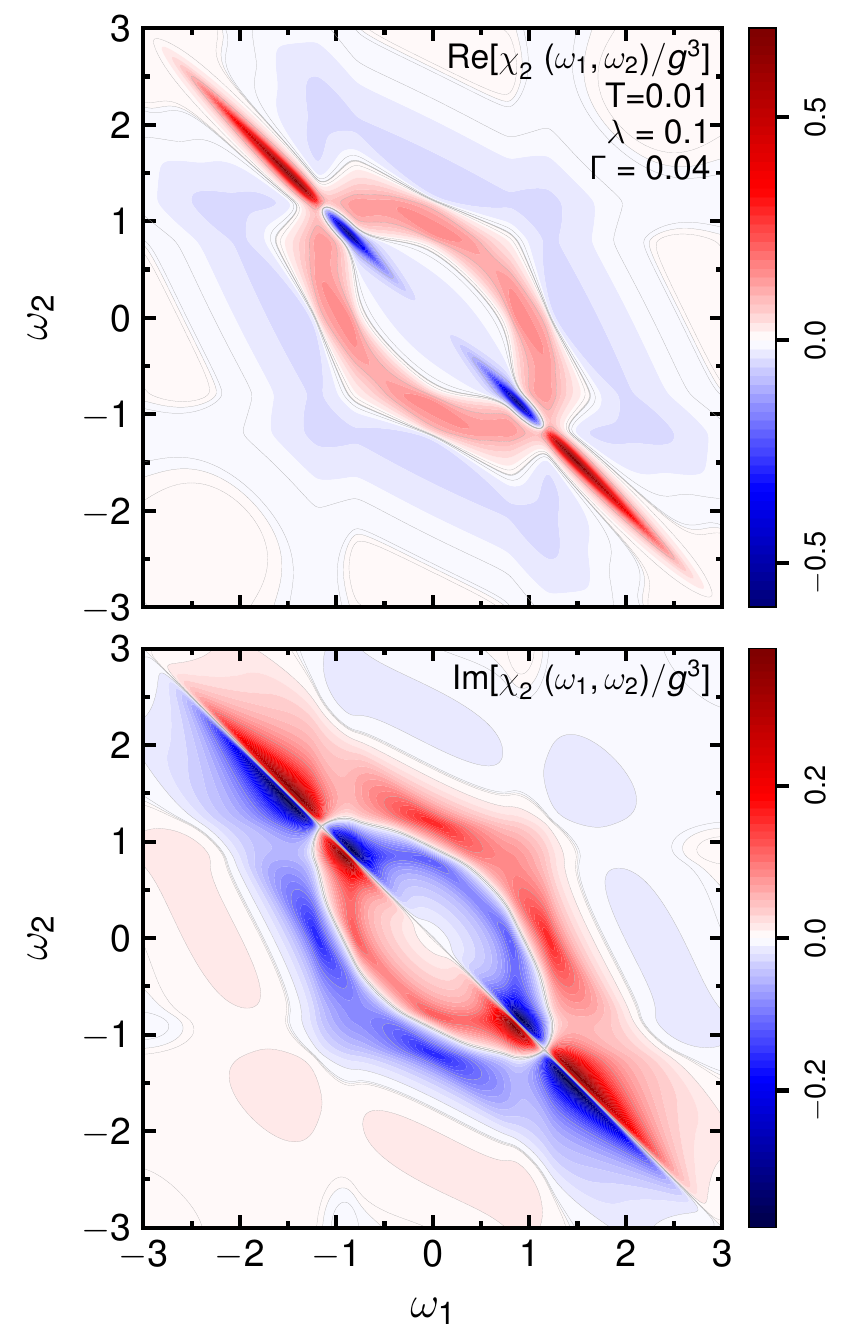}\vskip -.3cm\caption{Contours of the real and imaginary part of the 2DCS susceptibility
$\chi_{2}(\omega_{1}{,}\omega_{2})$ in the homogeneous gauge state
at fixed static field $\lambda=-gE_{dc}$. Linear system size $L{=}400$,
energies in units of $J$. \label{fig:o2cont}}
\end{figure}

Anticipating the latter integration over $\mathbf{k}$ and leaving
the polarization matrix elements aside, we now analyze the contributions
to Eq. (\ref{eq:c2s}) arising from the pole structures of Eq. (\ref{eq:tlchi}).
They comprise two kinds of singularities. The first kind occurs if
only one of the differences of squares in the denominator vanishes.
In the real $\omega_{1,2}$-plane this produces simple $\delta$-functions.
Integrating them over $\epsilon\rightarrow\epsilon_{\mathbf{k}}$
as in Eq. (\ref{eq:c2s}) leads to finite and smooth contributions
in the thermodynamic limit. The second kind of singularity occurs
if two of the differences of squares in the denominator vanish simultaneously.
This comprises four types of poles: (i) $z_{1}=z_{2}=\pm2\epsilon$,
(ii) $z_{1}=-z_{2}=\pm2\epsilon$, (iii) $z_{1(2)}=\pm2\epsilon$,
$z_{2(1)}=0$, and (iv) $z_{1(2)}=-z_{2(1)}/2=\pm2\epsilon$. Note
that the singularity of type (i) marks a resonance of $\langle\Delta P\rangle(t)$
while keeping an output frequency $\omega_{1}+\omega_{2}=2\omega_{1}$
which is twice the input. I.e., this is a resonant\emph{ second harmonic
generation} (SHG). Similarly, type (ii) corresponds to a resonant
response while keeping a DC output frequency $\omega_{1}+\omega_{2}=0$.
I.e., a resonant \emph{galvanoelectric effect} (GEE). The asymptotic
behavior of $\chi_{2}(z_{1}{,}z_{2})/(n|m|^{2})$ at one exemplary
pole for each of the types, (i)-(iv), is
\begin{equation}
\begin{array}{lc}
\mathrm{(i)}\,\omega{\equiv}\omega_{1}{=}\omega_{2}{\approx}2\epsilon, &
\frac{1}{2\epsilon}\frac{1}{\omega{-}2\epsilon{+}i\eta}\\
\mathrm{(ii)}\,\omega{\equiv}\omega_{1}{=}{-}\omega_{2}{\approx}2\epsilon, & \frac{1}{2i\eta}(\frac{1}{\omega{-}2\epsilon{-}i\eta}{-}\frac{1}{\omega{-}2\epsilon{+}i\eta})\\
\mathrm{(iii)}\,\omega{\equiv}\omega_{1}{\approx}2\epsilon,\,\omega_{2}{=}0, & \frac{1}{i\eta}(\frac{1}{\omega{-}2\epsilon{+}i\eta}{-}\frac{1}{\omega{-}2\epsilon{+}2i\eta})\\
\mathrm{(iv)}\,\omega{\equiv}\omega_{1}{=}{-}\frac{\omega_{2}}{2}{\approx}2\epsilon, & \frac{1}{12\epsilon}(\frac{1}{\omega{-}2\epsilon{+}i\eta}{-}\frac{1}{\omega{-}2\epsilon{-}2i\eta})
\end{array}\label{eq:polestruc}
\end{equation}
Integrating expressions of type (i), (iii), or (iv) over $\epsilon\rightarrow\epsilon_{\mathbf{k}}$
yields finite and smooth contributions to Eq. (\ref{eq:c2s}) in the
thermodynamic limit. For type (iii) this stems from the term in the
()-brackets contributing only at $O(\eta)$. This leaves the type
(ii) resonance on the GEE line. Asymptotically, for $\eta\ll2\epsilon$,
it can also be written as
\begin{equation}
\chi_{2}(z_{1}{,}z_{2})=n|m|^{2}\,\frac{\pi}{\eta}\,\delta(\omega_{1}-2\epsilon)\,,\,\,\,\omega_{1}{=}{-}\omega_{2}{\approx}2\epsilon\,.\label{eq:geesing}
\end{equation}
In turn, remarkably, $\chi_{2}(z_{1}{,}z_{2})$ on the GEE line is
anomalously singular, apart from being strictly real. I.e., integration
of Eq. (\ref{eq:geesing}) over $\epsilon$, as in Eq. (\ref{eq:c2s}),
yields a response \emph{diverging} $\propto1/\eta$ as the causal
broadening approaches zero \citep{Floquet}. Similar conclusions have
also been drawn for the GEE in other contexts \citep{Parker2019,Sipe2000,Fei2020,Ishizuka2022,Raj2024}.
Hereafter, and in order to regularize this singularity, we follow
the latter works and replace the causal broadening $\eta$ by a \emph{physical
relaxation rate} $\Gamma$ which remains a free parameter. Since the
Kitaev magnet maps to free Dirac fermions, we assume that $\Gamma$
arises from many-body interactions beyond this model.

\begin{figure}[tb]
\centering{}\includegraphics[width=0.45\columnwidth]{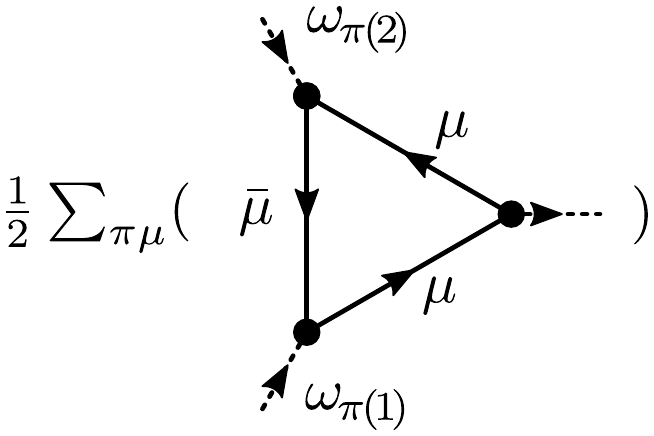}\caption{Diagrams contributing singularly to the GEE. $\pi$ refers to the
two permutations of input frequencies. $\mu{=}1{,}2$ refers to band
index and $\bar{1}{,}\bar{2}=2{,}1$. \label{fig:singdiags}}
\end{figure}

\subsection{Homogeneous flux state $T\lesssim T^{\star}$\label{subsec:Homogeneous-flux}}

To substantiate the preceding subsection, we now discuss several plots
of $\chi_{2}(z_{1}{,}z_{2})$. For that purpose we refrain from repeated
explicit display of the broadening within the arguments of $\chi_{2}$,
i.e., we set $\chi(\omega_{1}{,}\allowbreak\omega_{2})\equiv\allowbreak\chi_{2}(\omega_{1}+i\Gamma{,}\allowbreak\omega_{2}+i\Gamma)$.
In the homogeneous gauge state, the sole temperature dependence of
the 2DCS susceptibility is due to the Fermi function in Eq. (\ref{eq:c2s}).
For $T\ll J$ this is negligible. For numerical reasons we therefore
fix $T=0.01J\lesssim T^{\star}$ in this subsection. In Fig. \ref{fig:o2cont}
contours of the 2DCS susceptibility are shown for a typical set of
parameters, $\lambda$ and $\Gamma$. This plot clearly features a
main point of this work, associated with the analytic properties discussed
using the two-level model. Namely, $\chi(\omega_{1}{,}\omega_{2})$
is a smoothly varying function, except for a dominant antidiagonal
feature running along $\omega_{2}=-\omega_{1}$. This feature is related
exactly to the $1/\Gamma$-singular contribution which stems from
Eq. (\ref{eq:geesing}). As anticipated from that equation, the plot
also shows that $\chi_{2}(\omega_{1}{,}-\omega_{1})$ is purely real,
with
\begin{align}
\lefteqn{\chi_{2}(\omega_{1}{,}{-}\omega_{1})\simeq\frac{\pi g^{3}}{\Gamma}\tilde{\sum_{\mathbf{k}}}\left[(1{-}2f_{\mathbf{k}})\,p_{11}(\mathbf{k})\times\right.}\nonumber \\
 & \hphantom{aaaaaaaaaa}\left.|p_{12}(\mathbf{k})|^{2}(\delta(\omega+2\epsilon_{\mathbf{k}})+\delta(\omega-2\epsilon_{\mathbf{k}}))\right]\,.\label{eq:c2gee2}
\end{align}
for $\Gamma\ll J$. At this point it is obvious that the identification
of the causal broadening with a physical relaxation\emph{ }rate $\Gamma$
can be generalized readily to include momentum dependence by replacing
$\Gamma$ with $\Gamma_{\mathbf{k}}$, moved into the summation over
$\mathbf{k}$.

While Eq. (\ref{eq:c2gee2}) has already been made plausible via the
two-level model, we also mention that it is straightforward to show
that this singular behavior on the GEE line stems from 4 of the 16
graphs of Fig. \ref{fig:diags}, namely those depicted in Fig. \ref{fig:singdiags}.
Moreover, and while the introduction of $\Gamma$ seems rather \emph{ad-hoc},
it is also straightforward to show that introducing a phenomenological
one-particle selfenergy, $\Sigma_{\mu,\mathbf{k}}(z)=i\Gamma_{\mu,\mathbf{k}}\,\mathrm{sign}(\mathrm{Im}(z))$,
into the Dirac fermion Green's functions in Fig. \ref{fig:singdiags}
reproduces Eq. (\ref{eq:c2gee2}).

\begin{figure}[tb]
\centering{}\includegraphics[width=0.95\columnwidth]{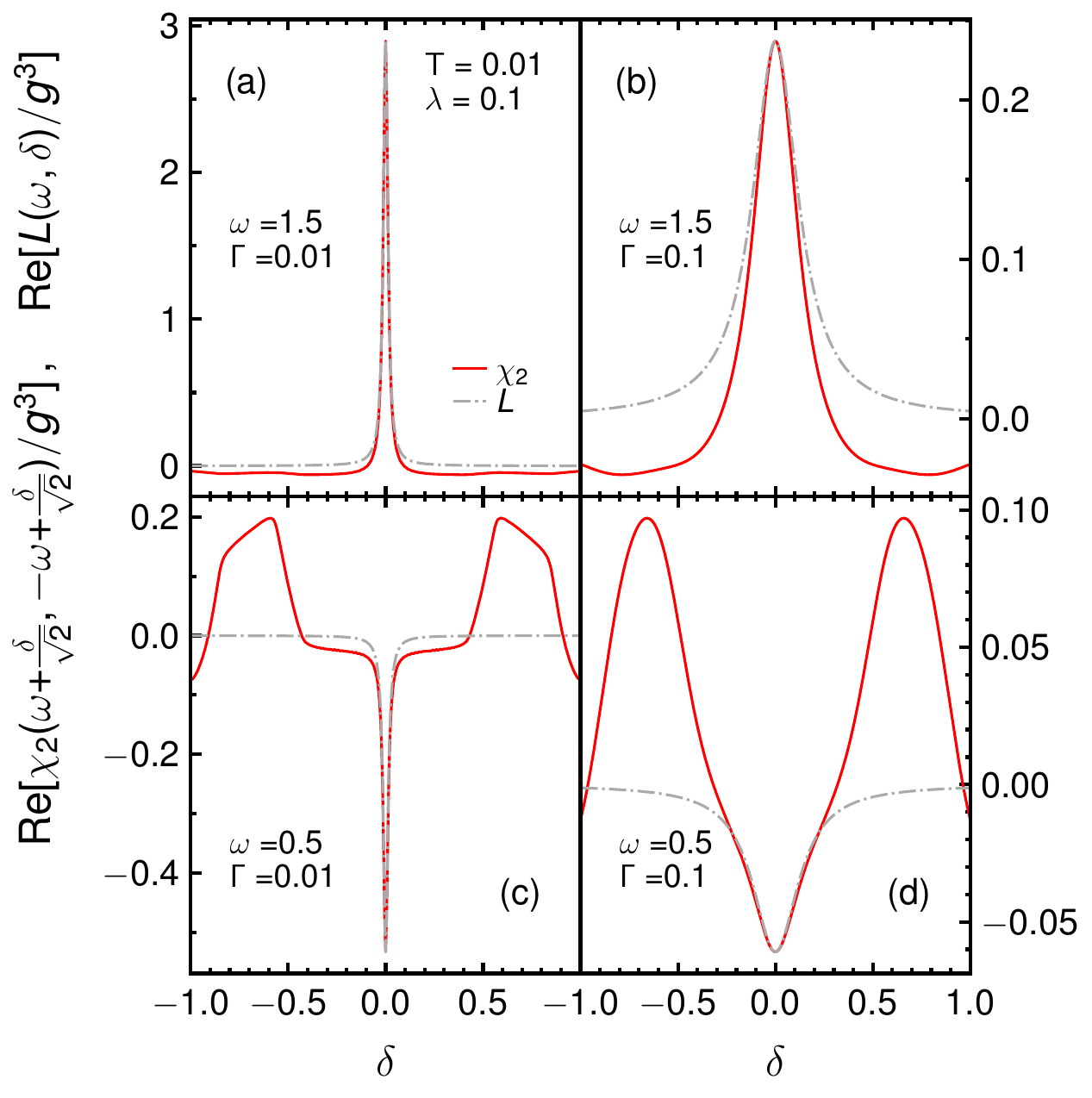}\vskip -.3cm\caption{Real part of 2DCS susceptibility $\chi_{2}(\omega_{1}{,}\omega_{2})$
in homogeneous gauge state, versus frequency $\delta$ perpendicular
to the GEE line (solid red) at fixed static field $\lambda=-gE_{dc}$,
compared to the asymptotic expression Eq. (\ref{eq:perpS}) (gray
dashed dotted), for two exemplary frequencies on the GEE line $\omega$
and two exemplary relaxation rates $\Gamma$. Linear system size $L{=}400$,
energies in units of $J$. \label{fig:perpgee}}
\end{figure}

Next, we clarify the asymptotic behavior of $\chi_{2}(\omega_{1},\omega_{2})$
in the vicinity of the GEE line by introducing coordinates $\omega$
and $\delta$, along and perpendicular to this line, $\omega_{1}=\omega+\delta/\sqrt{2}$
and $\omega_{2}=-\omega+\delta/\sqrt{2}$, respectively. Here, and
from the discussion of Eq. (\ref{eq:polestruc}) and for $\omega_{1}{\approx}{-}\omega_{2}{\approx}\pm2\epsilon$,
the poles of type $1/(((\omega_{1}+i\Gamma)\mp2\epsilon_{\mathbf{k}})((\omega_{2}+i\Gamma)\mp2\epsilon_{\mathbf{k}}))$
are relevant for $\omega_{1}\gtrless0$ in Eq. (\ref{eq:c2s}). Therefore,
asymptotically
\begin{align}
\chi_{2}(\omega_{1},\omega_{2}) & \sim\int d\epsilon\rho(\epsilon{,}\omega_{1{,}2})\frac{1}{\omega_{1}+2\epsilon+i\Gamma}\frac{1}{\omega_{2}-2\epsilon+i\Gamma}\nonumber \\
 & \sim\frac{i\sqrt{2}\,\Gamma\,\chi_{2}(\omega,-\omega)}{\delta+i\sqrt{2}\,\Gamma}\equiv L(\omega{,}\delta)\,,\label{eq:perpS}
\end{align}
where $\rho(\epsilon{,}\omega_{1{,}2})$ is a slowly varying function
resulting from the non-singular contributions of the sum over $\mathbf{k}$
in Eq. (\ref{eq:c2s}). To simplify, this is assumed to be constant
performing the $\epsilon$-integration. In Fig. \ref{fig:perpgee}
the asymptotic expression is compared with the actual 2DCS susceptibility
for two selected frequencies $\omega$. It shows reasonable agreement
close to the GEE line.

The preceding clarifies another central point of this paper. Namely,
tuning slightly off the GEE line, the real part of the 2DCS $O(2)$
low-frequency response displays a Lorentzian, the line width of which
corresponds to the one-particle relaxation rate. I.e., 2DCS spectroscopy
can be used to disentangle the continuum of fermion excitations, generated
by $P$, in order to access one-particle life-times.

\begin{figure}[tb]
\centering{}\includegraphics[width=0.95\columnwidth]{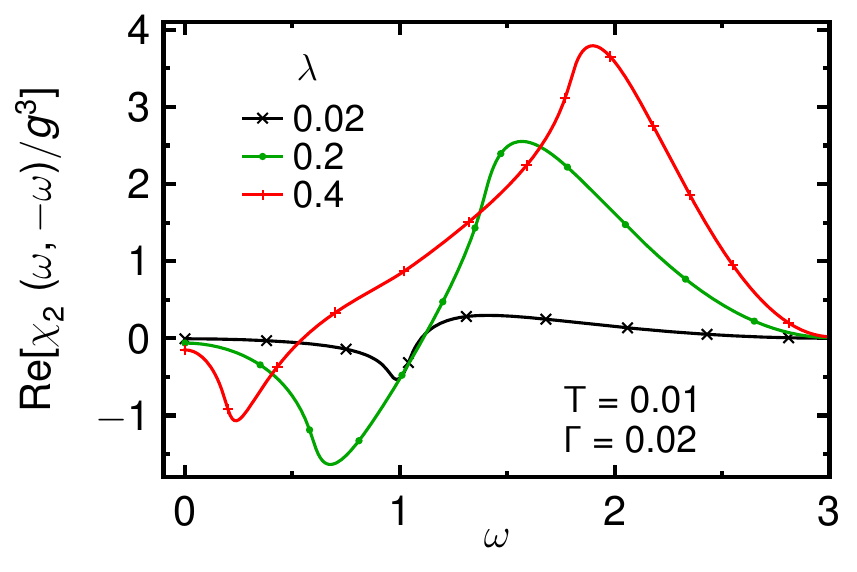}\caption{Real part of 2DCS susceptibility $\chi_{2}(\omega{,}-\omega)$ in
homogeneous gauge state versus $\omega$ along the GEE line for various
static fields $\lambda=-gE_{dc}$. Linear system size $L{=}400$,
energies in units of $J$. \label{fig:KvsEdcMark} }
\end{figure}

Cutting through the contour plots of Fig. \ref{fig:o2cont}, two directions
are of particular interest, i.e., $\omega_{1}=\omega_{2}=\omega$
and $\omega_{1}=-\omega_{2}=\omega$. The former refers to SHG and
has been discussed in ref. \citep{Krupnitska2023}. The latter is
the response function for the GEE. In Fig. \ref{fig:KvsEdcMark} this
is displayed for various DC fields. The figure also highlights the
relevance of explicitly breaking the $U$-symmetry, discussed in Sec.
\ref{sec:model}, in order to obtain a finite $O(2)$ 2DCS. I.e.,
for a vanishing DC field, $\lambda=0$, the susceptibility vanishes
and nonlinear response starts only at $O(3)$ of the electric field.

Finally, we mention that $\chi_{2}(z_{1},z_{2})$ shows a Fermi-blocking
effect from the factor of $(1-2f_{\mathbf{k}})$ in Eq. (\ref{eq:c2s}).
I.e., raising the temperature, the magnitude of the susceptibility
decreases globally due to the blocking of occupied states. However,
as mentioned already, within the temperature range $T\lesssim T^{\star}\ll J$
of the homogeneous state, all temperature variations are very small
only and we refrain from plotting this.

\begin{figure}[tb]
\centering{}\includegraphics[width=0.95\columnwidth]{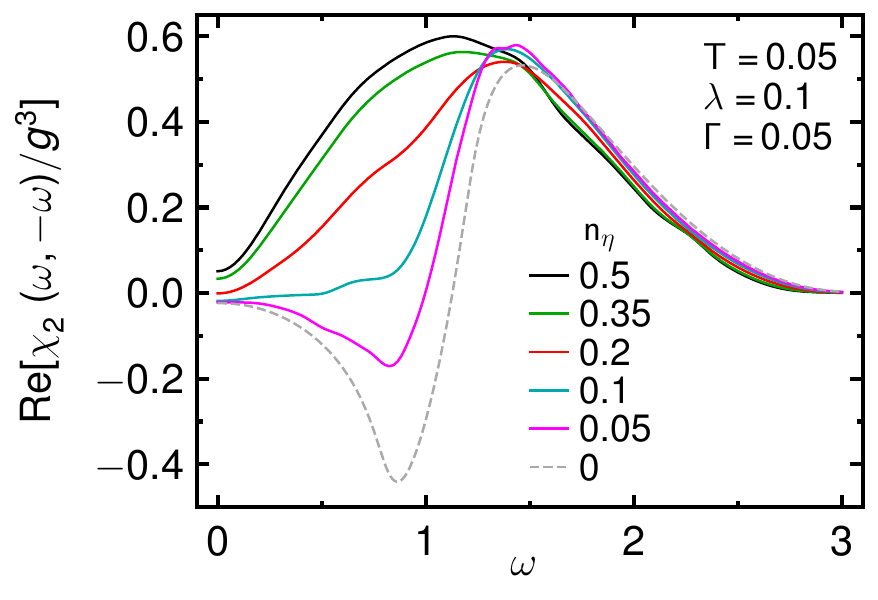}\caption{Solid colors: Real part of 2DCS susceptibility versus $\omega$ along
the GEE line, for various flipped gauge link densities $n_{\eta}$
at fixed temperature $T=0.05$ and static field $\lambda=-gE_{dc}$.
Linear system size $L=30$, number of random realizations $62$. Dashed:
Same, however, in homogeneous gauge state with linear system size
$L=400$. Energies in units of $J$. \label{fig:revsfdens}}
\end{figure}

\subsection{Random flux state $T\gtrsim T^{\star}$\label{subsec:Random-flux}}

In this subsection we discuss 2DCS in a state with randomly flipped
gauge links which models the thermal occupation of visons \citep{Metavitsiadis2020,Metavitsiadis2022,Metavitsiadis2017,Pidatella2019,Metavitsiadis2017a,Krupnitska2023}.
We begin with an approximate description of the flux proliferation
crossover of the 2DCS on the GEE line. While for $T>0.05J$, flux
in the Kitaev model is essentially random, freezing-out the vison
excitations in the range of $0.01\lesssim T/J\lesssim0.05$ cannot
be treated exactly for three-point correlation functions like $\chi_{2}(\omega_{1},\text{\ensuremath{\omega}}_{2})$
at present. Yet, a phenomenological understanding of the flux proliferation
can be reached by fixing the fermion temperature at $T\sim0.05$ and
subsequently varying the average density of randomly flipped links
$0<n_{\eta}<1/2$, thereby approximating the crossover between the
homogeneous and the random gauge state. The effects of this are depicted
in Fig. \ref{fig:revsfdens}. First, the figure provides a satisfying
consistency check, since $\chi_{2}(\omega{,}{-}\omega)$ obtained
from the $\mathbf{r}$-space formalism encoded in Eq. (\ref{eq:c2srnd})
smoothly evolves into that from the $\mathbf{k}$-space formulation
Eq. (\ref{eq:c2s}), as the density of flipped gauge links approaches
zero. Second, the figure highlights another main point of this work,
namely that vison excitations have a drastic impact on the susceptibility,
suppressing an oscillatory behavior which is present without visons.
In turn, not only the fingerprints of the fermionic fractional quasiparticles
are encoded in the 2DCS response but also the second kind of fractional
quasiparticles, i.e., visons impact the susceptibility.

\begin{figure}[tb]
\centering{}\includegraphics[width=0.95\columnwidth]{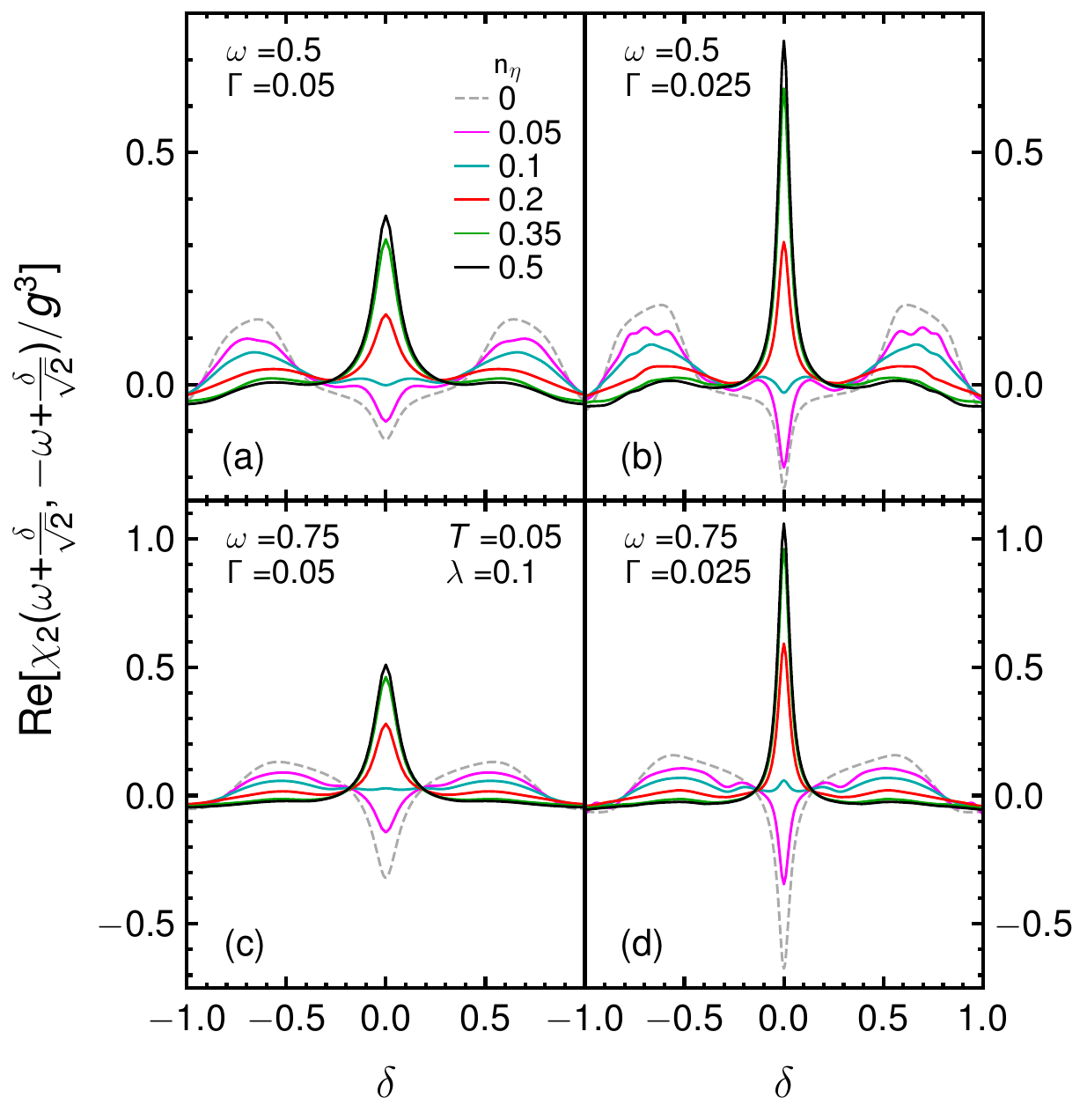}\vskip -.3cm\caption{Real part of 2DCS susceptibility $\chi_{2}(\omega_{1}{,}\omega_{2})$
versus frequency $\delta$ perpendicular to the GEE line (solid colors),
for various flipped gauge link densities $n_{\eta}$, at fixed temperature
$T=0.05$ and static field $\lambda=-gE_{dc}$, compared to the asymptotic
expression Eq. (\ref{eq:perpS}) (gray dashed), for two exemplary
frequencies $\omega$ on the GEE line and for two exemplary damping
rates $\Gamma$. Linear system size for randomized gauge $L{=}30$
with $62$ random realizations, for homogeneous state $L{=}400$.
Energies in units of $J$.\label{fig:TvarAtTstarPERP}}
\end{figure}

Next we consider the vicinity of the GEE line in the presence of gauge
excitations. As discussed in the previous subsection, in the homogeneous
gauge state, the width of $\mathrm{Re}[\chi_{2}(\omega_{1},\omega_{2})]$
allows to read off the one-particle relaxation rate $\Gamma$. Since
gauge excitations leave the Kitaev model a two-band one-particle Hamiltonian,
albeit with randomized energies and polarization operator matrix elements,
no additional relaxation channel is introduced by visons. Therefore
and as another main message, while on a global scale of the 2D frequency
plane, visons will strongly modify the susceptibility, see Fig. \ref{fig:revsfdens},
we expect the width of cuts perpendicular to GEE line, and in its
vicinity, to remain insensitive to gauge excitations. Exactly this
can be observed in Fig. \ref{fig:TvarAtTstarPERP}. For two exemplary
frequencies $\omega$ and damping rates $\Gamma$, the real part of
the 2DCS susceptibility is not only shown in the homogeneous and the
completely random gauge state, but for completeness also at flipped
gauge link densities less than $1/2$. Obviously, visons do impact
$\mathrm{Re}[\chi_{2}(\omega_{1},\omega_{2})]$, even so far as to
change its sign, similar to the low-frequency behavior in Fig. \ref{fig:revsfdens}.
However, the width of the central peak at $\delta\approx0$ remains
essentially unaffected. Therefore, it is tempting to speculate that
any $T$-dependence of this GEE line-width may provide additional
information on many-body interactions beyond the Kitaev model for
all temperatures.

\begin{figure}[tb]
\centering{}\includegraphics[width=0.95\columnwidth]{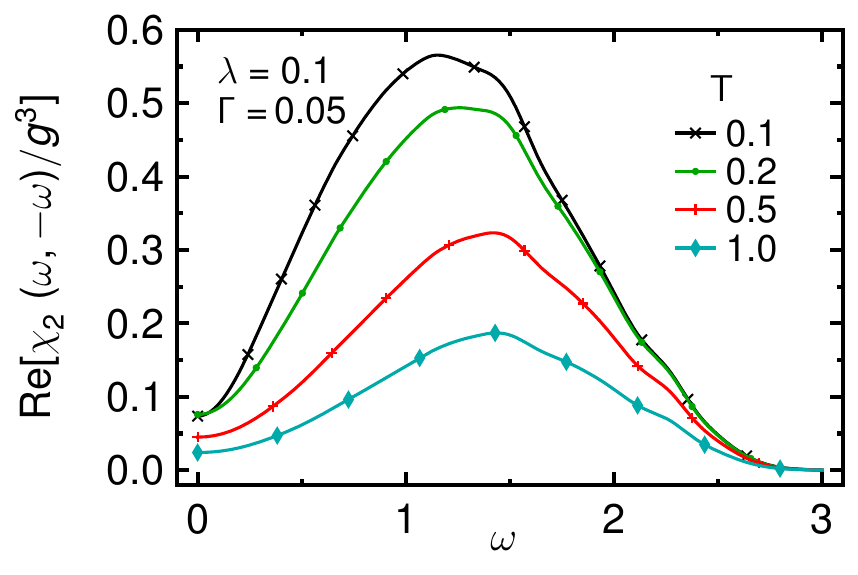}\caption{Real part of 2DCS susceptibility in random gauge state versus $\omega$
along the GEE line, for various temperatures $T$ at fixed static
field $\lambda=-gE_{dc}$. Linear system size $L=30$, number of random
realizations $62$, energies in units of $J$.\label{fig:rnd05vsT}}
\end{figure}

Finally, we show Fermi blocking in $\chi_{2}(\omega_{1},\omega_{2})$
for $T\gtrsim T^{\star}$, i.e., in the fully random gauge state.
This is displayed in Fig. \ref{fig:rnd05vsT} along the GEE line for
$0.1J\leq T\leq J$. Clearly, the main feature of this plot is a continuous
suppression of the 2DCS susceptibility upon increasing the temperature
which results from fermionic states becoming blocked once they are
thermally populated. This can be viewed as an evidence for the fermionic
statistics of fractional excitations. It is completely in line with
the same effect seen in SHG in ref. \citep{Krupnitska2023} and also
reported for other spectroscopies in Kitaev magnets, including light
scattering \citep{Sandilands2015,Nasu2016,Wulferding2020,Halasz2016},
and phonon dynamics \citep{Metavitsiadis2020,Metavitsiadis2022,Ye2020,Feng2021,Feng2022,Li_diff_2021,Hauspurg2023}.

\section{Summary\label{sec:Summary}}

In conclusion, we have studied electric field induced 2DCS in a Kitaev
magnet at finite temperature. While the 2DCS susceptibility is set
solely by the fermionic excitations and their coupling to the laser
field, the fermionic spectrum is strongly modified by thermally excited
visons. In turn, we find the 2DCS response to vary with temperature
not only via Fermi statistics, but, far more importantly, also via
intrinsic gauge randomness versus temperature. Strikingly, the 2DCS
susceptibility displays a strong antidiagonal GEE singularity, which
needs to be cut off by one-particle relaxation rates beyond the plain
Kitaev model. These relaxation rates can be extracted from the 2DCS
response perpendicular to the antidiagonal, which is robust against
the gauge disorder at elevated temperatures. It is tempting to suggest
that 2DCS experiments may therefore extract fractional quasiparticle
lifetimes from a spectrum that is otherwise a featureless superposition
of multi-particle excitations.

Our study has a number of loose ends that could be followed in future
work. First, the combination of the Kitaev magnet and the coupling
to electric fields via exchange-striction is motivated primarily by
its simplicity. Other, more realistic couplings should be analyzed.
It seems reasonable to speculate that gross features of the present
work are robust, if more general dipole operators remain to mediate
inter- and intraband transitions within the fermion bands. Second,
the need for cutting off the singularity on the GEE line by a scattering
rate, and apart from phenomenology, opens up a playground for interacting
theories beyond the bare Kitaev model. This pertains to one-particle
renormalizations, as well as to vertex corrections of the three-point
2DCS susceptibility. Interestingly, within a completely different
context \citep{Kappl2023}, such directions have been pursued for
other models recently. Next, the present work has focused on the lowest
order response, explicitly breaking symmetries via the DC field. Calculations
should however also be performed for higher order response. Finally,
electric field induced 2DCS should also be considered for quantum
spin systems other than the Kitaev magnet.\vspace{5mm}

\begin{acknowledgments}
Fruitful discussions with R. Valent\'{i}, M. M\"oller, and D. Kaib
are gratefully acknowledged. We thank A. Schwenke for critical reading
of the manuscript. Research of W.B. was supported in part by the DFG
through Project A02 of SFB 1143 (project-id 247310070) and by grant
NSF PHY-2309135 to the Kavli Institute for Theoretical Physics (KITP).
W.B. acknowledges kind hospitality of the PSM, Dresden.
\end{acknowledgments}

\appendix

\section{Quasiparticles of homogeneous state\label{sec:UTrafo}}

The unitary transformation $\mathbf{u}(\mathbf{k})$ to quasiparticles
used in Sec. \ref{subsec:LTsusc} and detailed in several refs., e.g.,
\citep{Metavitsiadis2020} reads 
\begin{align}
\left[\begin{array}{c}
c_{{\bf k}}\\
a_{{\bf k}}
\end{array}\right] & =\left[\begin{array}{cc}
u_{11}({\bf k}) & u_{12}({\bf k})\\
u_{21}({\bf k}) & u_{22}({\bf k})
\end{array}\right]\left[\begin{array}{c}
d_{1{\bf k}}\\
d_{2{\bf k}}
\end{array}\right]\label{eq:4}\\
u_{11}({\bf k}) & =-u_{12}({\bf k})=\frac{i\sum_{\alpha}e^{-i{\bf k}\cdot{\bf r}_{\alpha}}}{2^{3/2}\epsilon_{{\bf k}}}\nonumber \\
u_{21}({\bf k}) & =u_{22}({\bf k})=\frac{1}{\sqrt{2}}\,,\nonumber 
\end{align}
where ${\bf k}=x\,{\bf G}_{1}+y\,{\bf G}_{2}$ with $x,y\in[0,2\pi[$
and ${\bf G}_{1[2]}=(1,-\frac{1}{\sqrt{3}})\,[(0,\frac{2}{\sqrt{3}})]$
is the reciprocal basis of the triangular lattice with the direct
basis listed in Fig. \ref{fig:model}. From this $\mathbf{u}(\mathbf{k})$,
a quasiparticle energy of $\epsilon_{{\bf k}}=J[3+2\lambda^{2}+\allowbreak2(1-\lambda^{2})\cos(x)+\allowbreak2(1-\lambda)\cos(x-y)+\allowbreak2(1+\lambda)\cos(y)]^{1/2}/2$
is obtained in terms of the reciprocal coordinates, as well as the
matrix elements of the dipole operator $p_{11}(\mathbf{k})=-p_{22}(\mathbf{k})=(\cos(y)-\allowbreak\cos(x-y)+\allowbreak2\lambda(1-\allowbreak\cos(x)))/\allowbreak(4\epsilon_{{\bf k}})$
and $p_{12}(\mathbf{k})=p_{21}^{\star}(\mathbf{k})=-i(\sin(x-\allowbreak y)+\allowbreak2\sin(x)+\allowbreak\sin(y))/\allowbreak(4\epsilon_{{\bf k}})$,
also expressed in reciprocal coordinates.


\begin{thebibliography}{99}
\bibitem{RndSec}The term ``random flux state'' is used to imply
an average over Hilbert space sectors with fixed, random values of
$\text{\ensuremath{\eta_{\mathbf{l}}}}$.
\bibitem{Floquet}Analysis of an exponentially ramped two-level Floquet
Hamiltonian with a driving polarization as in this work, yields a
resonant GEE, which at second order is fully consistent with Eq. (\ref{eq:geesing})
\end{thebibliography}
\end{document}